\documentclass[journal]{IEEEtran}
\usepackage{amsmath}
\usepackage{amsthm}
\usepackage{amssymb}
\usepackage{amsfonts}
\usepackage[noend]{algpseudocode}
\usepackage{algorithmicx,algorithm}
\usepackage{cite}
\usepackage[dvipsnames]{xcolor}
\usepackage{subfig}
\usepackage{graphicx}
\usepackage{soul}
\usepackage{mathtools}
\usepackage{hyperref}
\usepackage{xcolor}
\usepackage{balance}
\pdfoutput=1

\graphicspath{{figs/}}

\ifodd 1

\newcommand{\CommentWong}[1]{\textcolor[rgb]{1,0,0}{[CW: #1]}}

\newcommand{\CommentRitesh}[1]{\textcolor[rgb]{0,0,1}{[Ritesh: #1]}}
\newcommand{\EditRitesh}[1]{\textcolor[rgb]{0,0,1}{#1}}
\else

\newcommand{\CommentWong}[1]{}

\newcommand{\CommentRitesh}[1]{}
\newcommand{\EditRitesh}[1]{}
\fi

\def\E{\mathbb{E}}
\def\x{\boldsymbol{x}}
\def\y{\boldsymbol{y}}
\def\q{\boldsymbol{q}}
\def\v{\boldsymbol{v}}
\def\V{\boldsymbol{V}}
\def\w{\boldsymbol{w}}
\def\A{\boldsymbol{A}}

\def\h{\boldsymbol{h}}
\def\s{\boldsymbol{s}}
\def\F{{\mathcal{F}}}

\def\F{\mathcal{F}}
\def\M{\mathcal{M}}
\def\Fviral{\F \text{ viral}}
\newcommand{\normaldensity}[3]{\mathcal{N}({#1}; {#2}, {#3})}

\renewcommand{\textcolor}[2]{{#2}}

\def\pif{\pi_{\text{vf}}}

\def\pih{\pi_{\text{vf}}}
\def\pip{\pi_{\text{ind}}}

\DeclarePairedDelimiterX{\norm}[1]{\lVert}{\rVert}{#1}

\renewcommand{\baselinestretch}{1.05}
\renewcommand{\Pr}{\operatorname{\mathbb{P}}}

\begin{document}

\title{Group Testing with Side Information\\
via Generalized Approximate Message Passing}

\author{Shu-Jie~Cao,
Ritesh~Goenka,
Chau-Wai~Wong,~\IEEEmembership{Member,~IEEE,}
Ajit~Rajwade,~\IEEEmembership{Senior Member,~IEEE,}
Dror~Baron,~\IEEEmembership{Senior Member,~IEEE}%
\thanks{This work was supported in part by the US National Science Foundation under Grant ECCS-2030430, and in part by India's Science and Engineering Research Board (SERB) Matrics Grant \#10013890, IITB-WRCB (Wadhwani Center for Bioengineering) Grant \#DONWR04-002, and DST-Rakshak Grant \#DST0000-005. A subset of this work was presented at the 2021 IEEE International Conference on Acoustics, Speech and Signal Processing (ICASSP) [DOI: 10.1109/ICASSP39728.2021.9414034]~\cite{Goenka2021}.}
\thanks{Shu-Jie Cao is with the Department of Electrical and Computer Engineering, Northwestern University, Evanston, IL 60208 USA (e-mail:
shujie.cao1023@gmail.com). Part of this work was conducted when Ms. Cao was with ShanghaiTech University, Shanghai 201210, China.}
\thanks{Ritesh~Goenka is with the Department of Mathematics, the University of British Columbia, Vancouver, BC V6T 1Z4, Canada (email: goenkaritesh12@gmail.com). Part of this work was conducted when Mr. Goenka was with IIT Bombay, Mumbai, MH 400076 India.}
\thanks{Chau-Wai Wong and Dror Baron are with the Department of Electrical and Computer Engineering, NC State University, Raleigh, NC 27695 USA (e-mail: chauwai.wong@ncsu.edu; barondror@ncsu.edu).}
\thanks{Ajit~Rajwade is with the Department of Computer Science and Engineering, IIT Bombay, Mumbai, MH 400076, India (e-mail: ajitvr@cse.iitb.ac.in).}
}

\maketitle

\begin{abstract}
Group testing can help maintain a widespread testing program using fewer resources amid a pandemic.
In a group testing setup, we are given $n$ samples, one per individual.
Each individual is either infected or uninfected. 
These samples are arranged into $m < n$ pooled samples, where each pool is obtained by mixing a subset of the $n$ individual samples. 
Infected individuals are then identified using a group testing algorithm.
In this paper, we incorporate side information~(SI) collected from contact tracing~(CT) into nonadaptive/single-stage group testing algorithms.
We generate different types of possible CT SI data by incorporating different possible characteristics of the spread of disease.
These data are fed into a group testing framework based on generalized approximate message passing (GAMP).
Numerical results show that our GAMP-based algorithms provide improved accuracy.
\end{abstract}


\IEEEpeerreviewmaketitle

\section{Introduction}
\label{sec:intro}
Widespread testing is a broadly used epidemiological tool for combating the COVID-19 pandemic.
Samples are typically collected from nasal or oropharyngeal swabs and then processed by a reverse transcription polymerase chain reaction~(RT-PCR) machine.
However, widespread testing may be hindered by supply chain constraints and long testing times, especially at the inception of a pandemic. 

Pooled or {\em group testing} has been suggested for improving testing efficiency \cite{Dorfman1943,aldridge2019group,Hogan2020,Abdelhamid2020,Zhu2020,Yi_arxiv,Ghosh2021,zhu2020paris,Heiderzadeh2020,Nikolopoulos2021a,Shental2020,lin2020comparisons,cohen2020multilevel,lin2021positively,Nikolopoulos2021b,ahn2021adaptive,arasli2021group}.
Group testing involves mixing a subset of $n$ individual samples into $m < n$ pools. The measurement process can be expressed as $\boldsymbol{y} = \mathfrak{N}(\boldsymbol{Ax})$,
where $\boldsymbol{x}$ is a vector that quantifies the health status of the $n$ individuals,
$\boldsymbol{A}$ is an $m \times n$ binary pooling matrix with $A_{ij} = 1$ if the $j$th individual contributes to the $i$th pool, else $A_{ij} = 0$,
$\boldsymbol{y}$ is a vector of $m$ noisy measurements or tests,
and $\mathfrak{N}$ represents a probabilistic noise model that relates the noiseless pooled results, $\boldsymbol{Ax}$, to 
$\boldsymbol{y}$.
Our work considers a {\em noisy binary measurement model}
used by Zhu et al.~\cite{Zhu2020}, 
where $\boldsymbol{x}$ is a binary {vector},
$\boldsymbol{w} = \boldsymbol{Ax}$
is an auxiliary (noiseless) vector,
and the measurement $y_i\in\{0,1\}$ depends probabilistically on $w_i$,
where $\Pr(y_i=1 \mid w_i=0)$ and $\Pr(y_i=0 \mid w_i>0)$ are probabilities of erroneous tests.

We wish to estimate $\boldsymbol{x}$ from $\boldsymbol{y}$ and $\boldsymbol{A}$.
We use single-stage {\em nonadaptive} algorithms as in \cite{Zhu2020, Ghosh2021}, rather than two-stage algorithms,
which employ a second stage of tests depending on results from the first stage, as in 
Heidarzadeh and Narayanan~\cite{Heiderzadeh2020} or the classical 
Dorfman approach~\cite{Dorfman1943}. 
The advantage of nonadaptive algorithms is that they reduce testing time, which is high for RT-PCR.

Algorithms that estimate $\boldsymbol{x}$ from $\boldsymbol{y}$ and $\boldsymbol{A}$~\cite{Ghosh2021, Shental2020} rely primarily on the {\em sparsity} of $\boldsymbol{x}$, which is a valid assumption for COVID-19 due to low prevalence rates \cite{Benatia2020}.
However, in addition to sparsity, the health 
status vector $\boldsymbol{x}$ contains plenty of 
statistical structure. For example,
Zhu et al.~\cite{Zhu2020} exploited probabilistic information such as the prevalence rate and structure in $\boldsymbol{x}$, and stated the potential benefits of using {\em side information} (SI).
Specific forms of SI include individuals' symptoms and family 
structure~\cite{zhu2020paris, Goenka2021}. Family structure refers to information regarding which individuals belong to the same family. Besides the conventional meaning of a family, within the context of disease spread, other types of family-like structure with significant contact between all individuals
could include a group of students sharing the same room in a hostel, security officers working regularly at the same checkpoint, or healthcare workers working in the same facility.
Finally, Nikolopoulos et al.~\cite{Nikolopoulos2021a,Nikolopoulos2021b} independently observed that community structure can improve the performance of group testing; 
these works focused on encoder design in conjunction with basic decoders.

\begin{figure*}[!t]
  \begin{tabular}{@{}cc@{}c@{}c@{}c@{}}
  \multicolumn{2}{l}{\underline{Sparsity:} \hspace{8mm} \underline{$2.12\%$}} & \underline{$3.98\%$} & \underline{$6.01\%$} & \underline{$8.86\%$}  \vspace{0mm} \\
  \rotatebox[origin=l]{90}{\hspace{14mm}\underline{\textbf{GAMP}}} & 
    \includegraphics[width=0.241\linewidth]{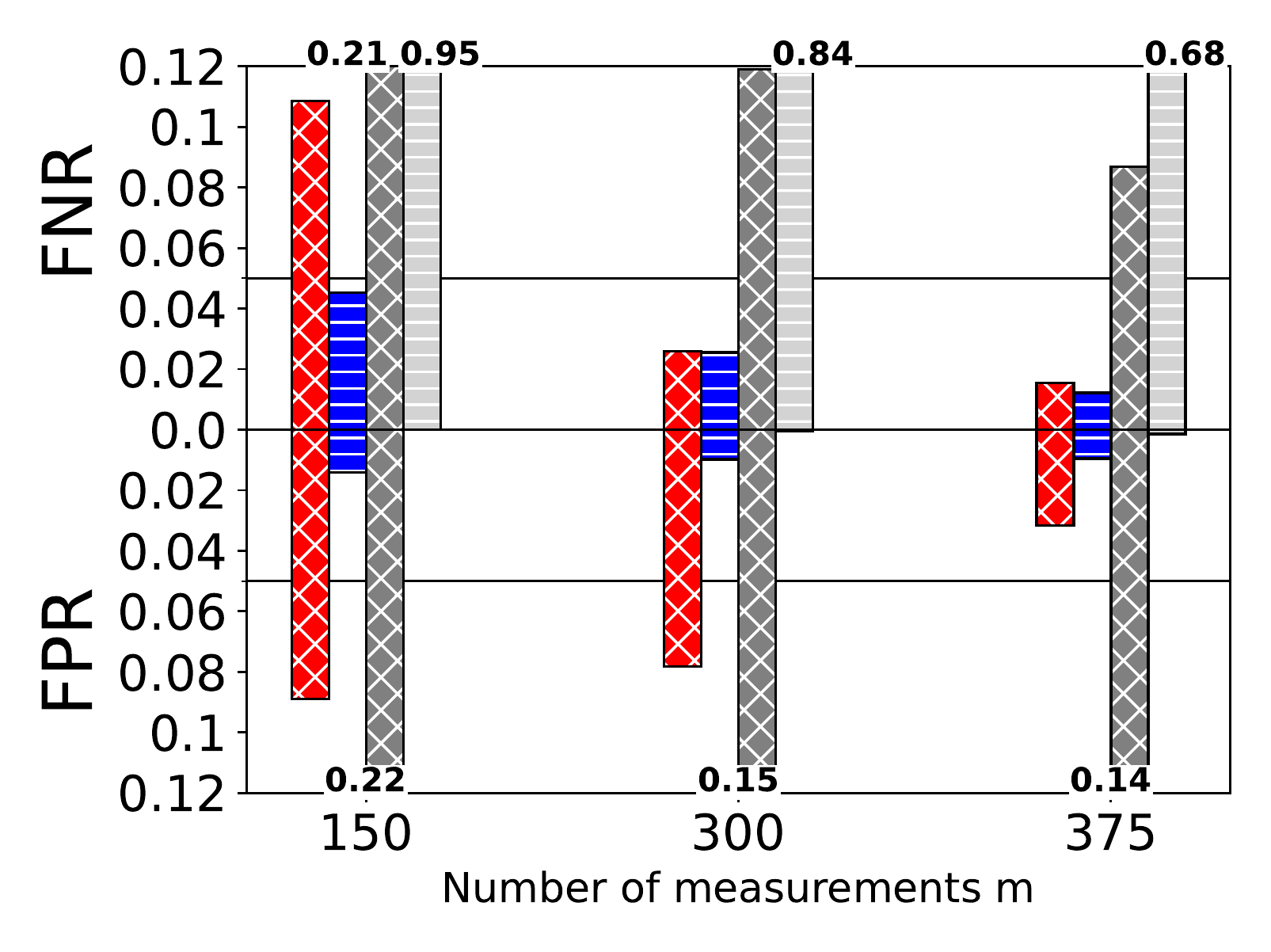} &
    \includegraphics[width=0.241\linewidth]{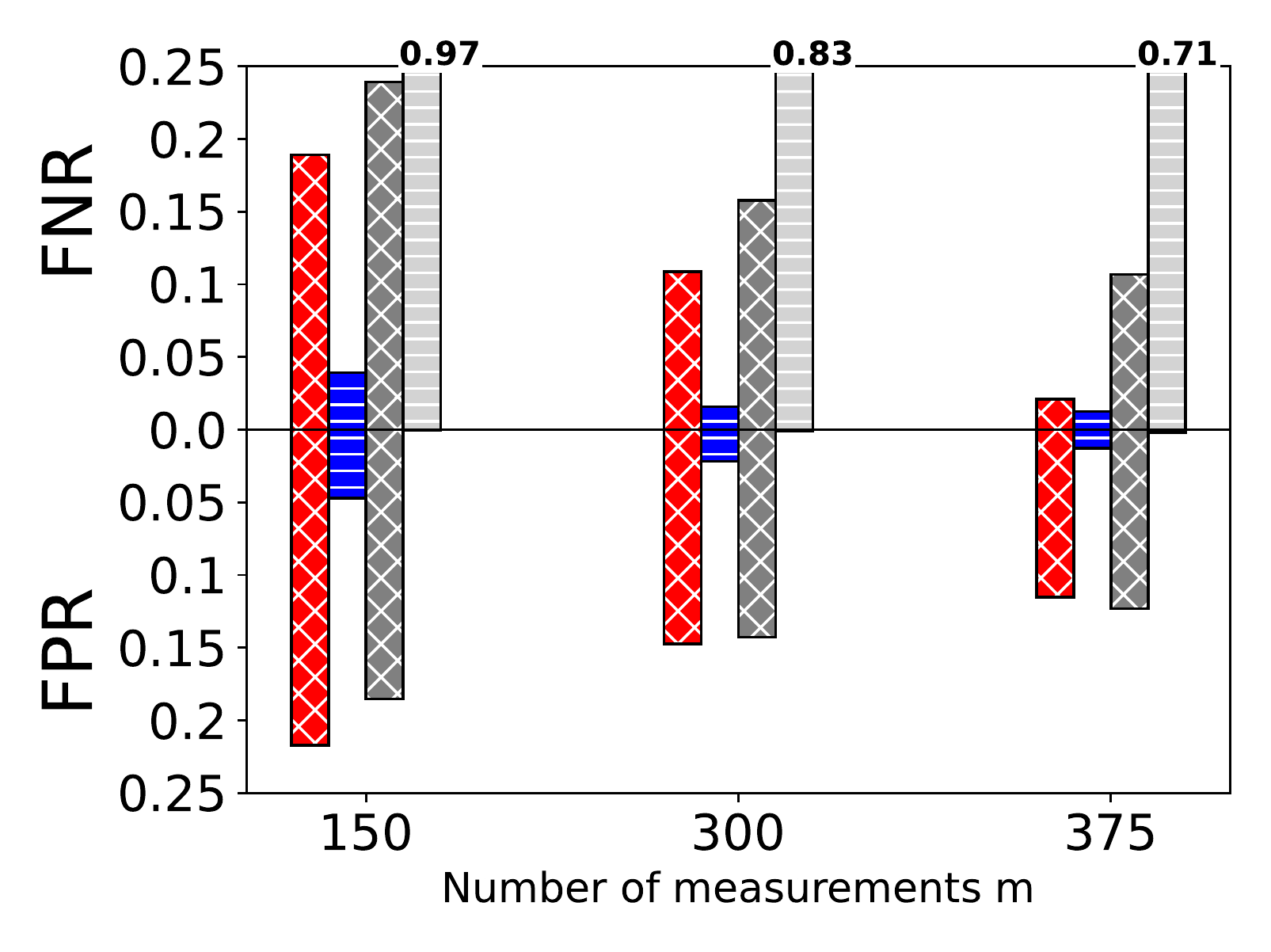} &
    \includegraphics[width=0.241\linewidth]{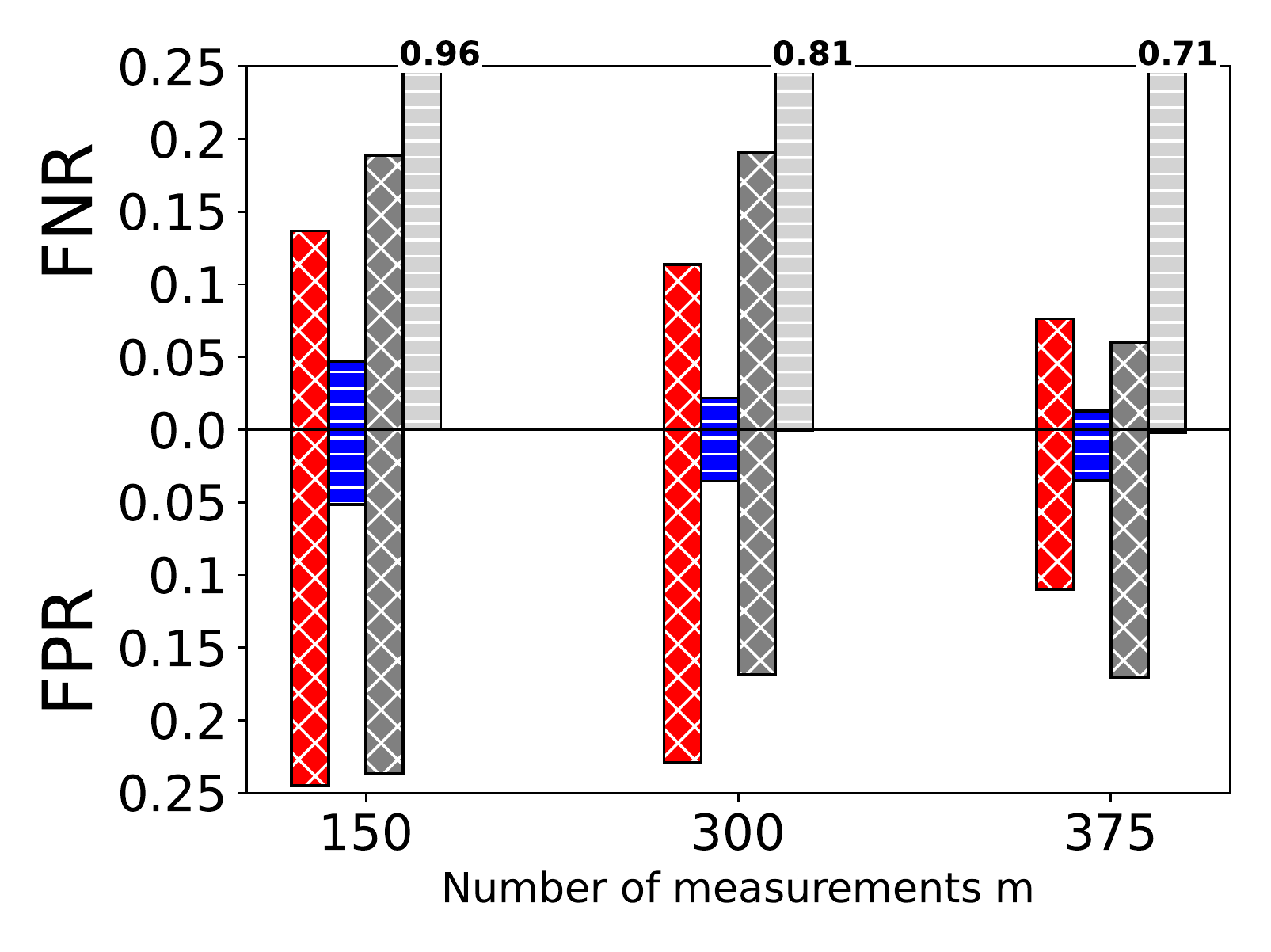} &
    \includegraphics[width=0.241\linewidth]{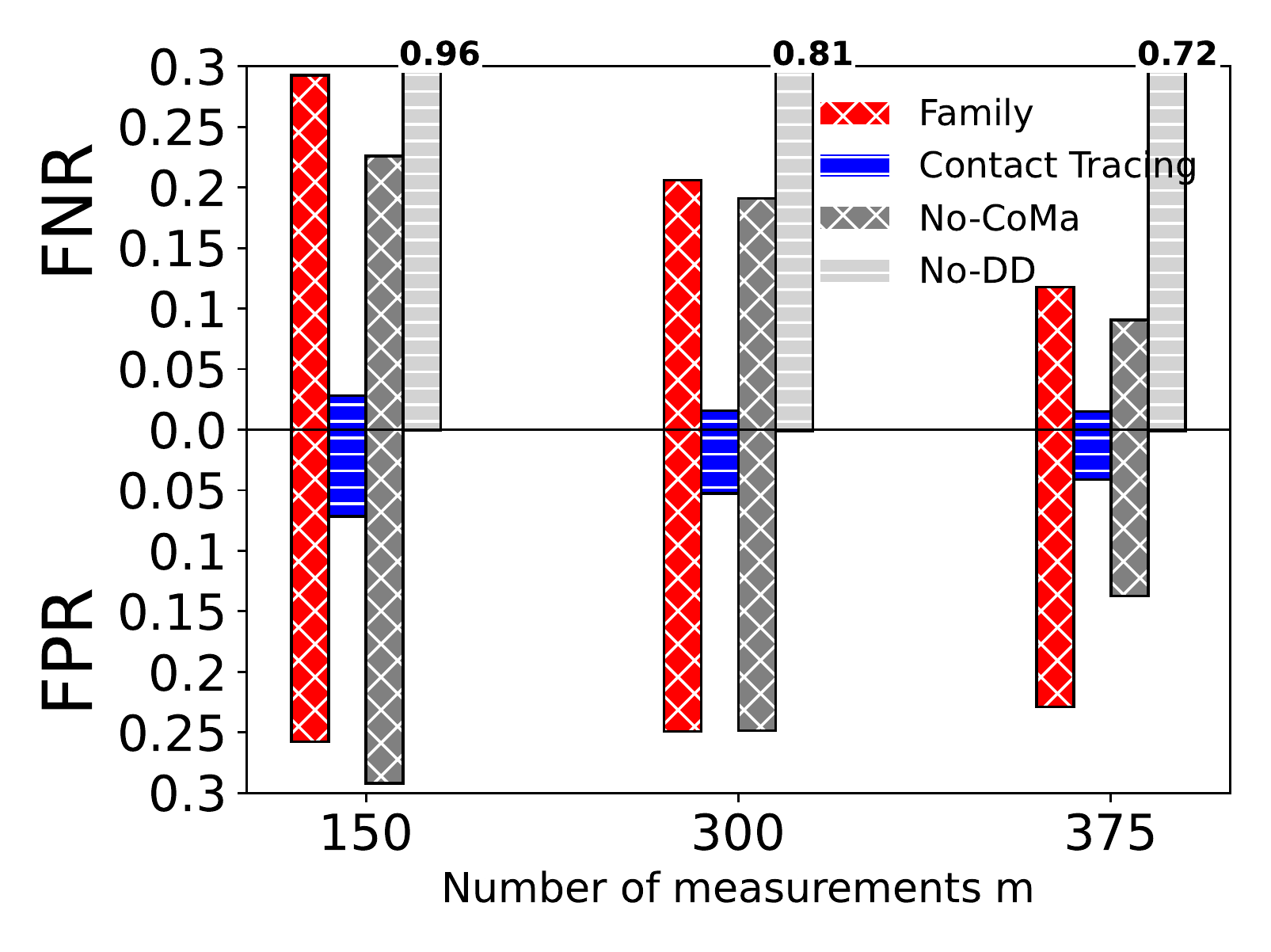} \vspace{-2mm} 
  \end{tabular}
  \caption{Performance of the proposed group testing method with binary noise at four averaged sparsity levels and three measurement levels for a population of $n = 1000$ individuals. Our algorithms use family and contact tracing information to obtain false negative rates~(FNR) and false positive rates~(FPR) that are more appealing than those obtained by the prior art \textcolor{red}{noisy-CoMa~\cite{CoMa} and noisy-DD~\cite{aldridge2019group}. See Sec.~\ref{subsec:results_M1} for more details. We also ran noisy-LiPo~\cite{CoMa} with the same datasets, and it has larger FNR than noisy-DD.}}
  \label{fig:res-m1-m2}
\end{figure*}

The focus of our work is on innovative decoder design.
Our algorithmic approach for the decoder uses approximate message passing (AMP)~\cite{donoho2009message}, 
which has been proven to achieve the best-possible estimation quality 
in the large system limit for certain matrices~\cite{celentano2022fundamental, dudeja2022universality, wang2022universality}.
Although the structured, binary, non-random measurement matrices that we use do not fall into this category, nonetheless, it seems plausible that our numerical results with these matrices are favorable.
Our AMP-based decoder is theoretically motivated, and
therefore expected to estimate $\boldsymbol{x}$ well,
in particular for large problem sizes.
{\em The overarching message of our paper is that 
using SI within AMP-based approaches can greatly improve group testing.}

{\em Contact tracing} (CT) information has been widely collected and used for
controlling a pandemic~\cite{cdc_contact_tracing}. 
Such information, including the duration of contact between pairs of individuals and measures of physical proximity, 
can be collected using modalities such as Bluetooth~\cite{Hekmati2020}, the global positioning system~\cite{Kleinman2020}, manual inquiries by social workers~\cite{Hohman2021, Ross2020}, and financial transaction data~\cite{CDC_CT,Kleinman2020}. 
Karimi et al.\cite{BPCG} used CT information to improve group testing using a loopy belief propagation (BP) algorithm.

In this paper, we show how to estimate $\boldsymbol{x}$ while utilizing CT SI.
Our contributions are twofold.
First, in Sec.~\ref{sec:algos},
we propose group testing decoding algorithms based on generalized approximate message passing~(GAMP)~\cite{rangan2011generalized},
and demonstrate how CT SI can be utilized.
Our numerical results
are presented in Sec.~\ref{sec:results};
a typical result appears in Fig.~\ref{fig:res-m1-m2}.
Our work uses more SI than other binary group testing algorithms ~\cite{Nikolopoulos2021a,Nikolopoulos2021b,ahn2021adaptive}
that only considered family/community
structure in binary group testing when designing 
their encoders.

Our second contribution is to the AMP~\cite{donoho2009message} community.
The prior art has considered vector denoisers within AMP, within GAMP~\cite{schniter2016vector}, and even SI-aided vector denoisers within AMP~\cite{baron2020mmwave}, including rigorous state evolution results~\cite{liu2019analysis, liu2022rigorous}. However, combining SI-aided vector denoisers within {\em generalized} AMP (GAMP) is new, to the best of our knowledge.
GAMP adds another layer of complexity beyond AMP,
hence combining these tools within GAMP could be
numerically unstable. 
Our favorable numerical results suggest that fusing these algorithmic tools together offers a good direction for
future work in SI-aided signal recovery.

We also mention that an extended online version of our 
work~\cite{Ritesh_G_journal} 
incorporates SI at the encoder. We demonstrate numerically that SI at the encoder provides little extra performance benefits over the utilization of SI only at the decoder. 

The rest of the paper is organized as follows.
In Sec.~\ref{sec:related_work}, we survey related work.
Our CT data is described in~Sec.~\ref{Sec:data_gen}.
Our main contributions appear in Sec.~\ref{sec:algos},
where we propose a GAMP-based group testing algorithm that exploits CT SI. 
Numerical results are presented in Sec.~\ref{sec:results}. 
We discuss our findings in Sec.~\ref{sec:discussion}
and conclude in Sec.~\ref{sec:conclusion}.

\section{Related Work}
\label{sec:related_work}

Side information (SI) can be derived from 
many data sources, including travel history, medical history, 
and symptoms. 
SI can be used to assign infection probabilities as prior probabilities to individuals, who can subsequently be classified into different risk levels.
Risk-aware group testing algorithms can then process
individuals based on their risk levels.

This risk-aware approach has proven to be effective. 
For example, Deckert et al.~\cite{Deckert2020} demonstrated a marked reduction in the number of tests required if pooling is performed on subjects with similar infection probabilities, as opposed to subjects with widely varying infection probabilities. 
Moreover, risk-specific algorithms can incorporate strategies to minimize the number of tests; McMahan et al.~\cite{McMahan2012} adopt such a hierarchical approach in conjunction with  classical Dorfman pooling~\cite{Dorfman1943}, identifying pool sizes that minimize the expected number of tests. 
In broadly similar work by Bilder et al.~\cite{Bilder2010},  individuals are sorted per their infection probabilities,  
and extra tests are assigned for high-risk individuals. 

There are also algorithms that make use of SI about family structures 
or contact tracing information. One recent example is 
our previous work with Zhu and Rivera~\cite{Zhu2020,zhu2020paris},
who used both individual infection probabilities derived from symptom SI and family information while performing the decoding. 

More recently, Nikolopolous et al. \cite{Nikolopoulos2021a, Nikolopoulos2021b} used SI about 
connected or overlapping communities to design good encoding matrices. The designed matrices boost the performance of basic decoders such as definite defectives (DD)~\cite{aldridge2019group}
or loopy belief propagation (LBP)~\cite{Baron2010}, in conjunction with a model for binary noise that resembles
our model.

The work that is likely closest to ours is 
Karimi et al.~\cite{BPCG}. Both works fully employ contract
tracing SI in the decoder. Additionally, both works consider
noisy measurements.

Ahn et al.\cite{ahn2021adaptive} and Arasli and Ulukus~\cite{arasli2021group} considered correlations represented by arbitrary graphs drawn from a stochastic block model, albeit for adaptive group testing, unlike the nonadaptive approach in this paper. Their approaches generalized the i.i.d. assumption often used in group testing. Such correlations could be derived from family or contact tracing information. Graphs generated from this model contain node clusters with dense connections within a cluster and sparse connections between clusters. Such structures are typical of many social network graphs. The authors in \cite{lin2021positively,Lendle2012} provided an analytical proof of cost reduction in Dorfman pooling if there is a positive correlation between the different samples being pooled. Moreover, Lin et al.~\cite{lin2020comparisons} also presented a hierarchical agglomerative algorithm for pooled testing in a social graph. We note that \cite{ahn2021adaptive,arasli2021group,lin2020comparisons,Lendle2012} do not consider measurement noise. 

Another approach was considered by Lau et al.~\cite{lau2022model}. They employ the Ising model to capture underlying structures and dependencies in the data. They solve the Ising model with quadratic and linear programming, resulting in performance improvements in group testing. Differently from our work, CT information is captured by edge strengths in the Ising model.

The recent approach by Gatta et al.~\cite{LaGatta2021} fits a graph neural network to data on COVID-19 infections, deaths, and recoveries in a certain geographic location. This approach allows the authors to estimate temporally varying parameters of differential equations associated with a susceptible, exposed, infectious, and recovered~(SEIR) model. 
While there are some similarities to our work, 
we have used parameters based on recent documents published by the World Health Organization~(WHO)~\cite{Deckert2020}, whereas Gatta et al.~\cite{LaGatta2021} use a data-driven approach to perform regression via a graph-based neural network. Note that a neural network requires a large amount of data for training, which would not be readily available at the beginning of a pandemic. In contrast, our approach uses only basic knowledge about the spread of disease.

\section{Data generation approach}
\label{Sec:data_gen}

This section summarizes the main points of a data generation system 
that we use to illustrate a possible GAMP-based group testing
decoding algorithm that uses CT.
Our data generation system resembles the SEIR approach~\cite{he2020seir,pandey2020seir,yang2020modified},
but is not a contribution per se.

Below, we only present features of the data generation system that are utilized for
algorithm design in Sec.~\ref{sec:algos}.
GAMP is more of an algorithmic framework than a single algorithm, and a specific GAMP approach is derived based on statistical information about the structure of the vector $\boldsymbol{x}$ and its dependence on CT information (Sec.~\ref{sec:algos}). 
For these reasons, our data generation system was created
merely to demonstrate the efficacy of this GAMP-based framework.

\begin{figure}[!t]
  \centering
  \centerline{\includegraphics[width=8cm]{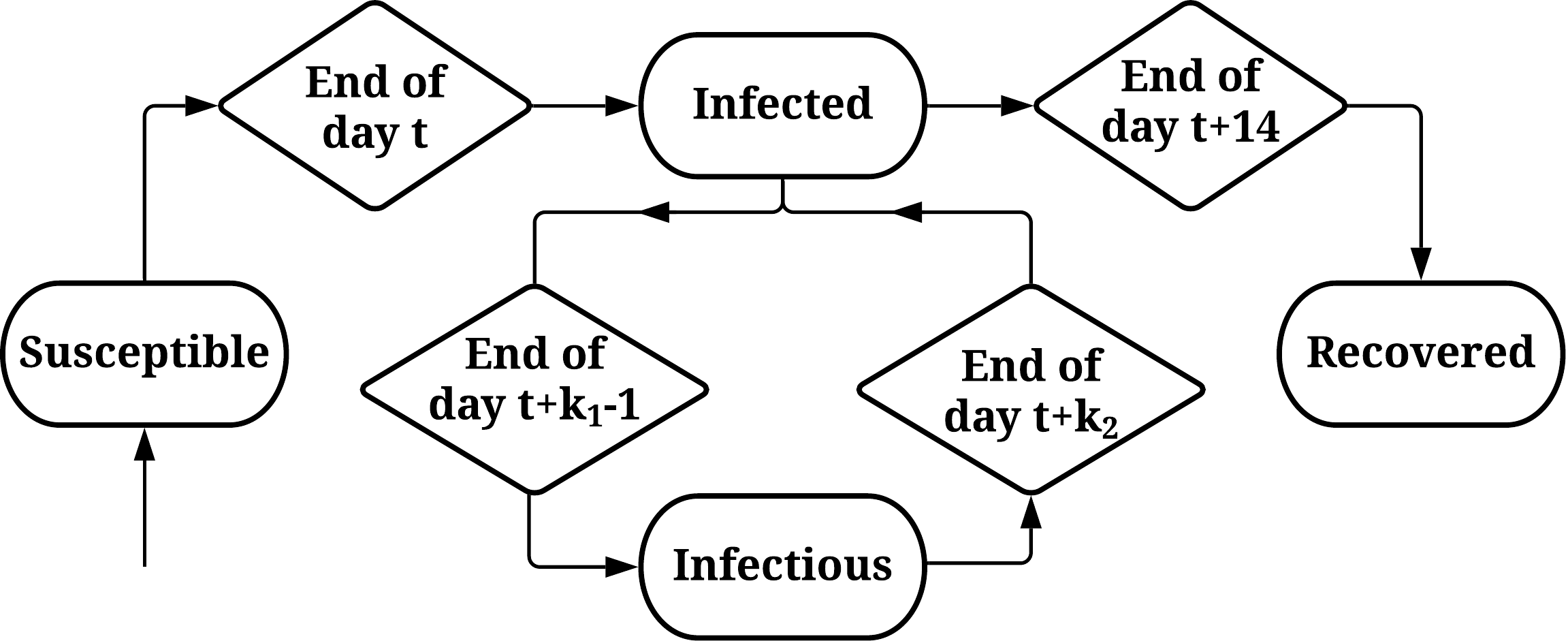}}
  \caption{Transitions between susceptible, infected, 
  infectious, and recovered per-node states,
  where each node represents an individual. 
  A node is infectious between days $k_1$ and $k_2$ (inclusive) after getting infected, where we set $(k_1,k_2) = (3,7)$.}
\label{fig:state-transition-diagram}
 \end{figure}

We model $n$ individuals using a dynamic time-varying graphical model, where for each day dynamic undirected edges represent recent
contacts between pairs of nodes, where nodes stand for individuals.
Each node can be in one of four
states, susceptible, infected, infectious, and recovered; for transitions between states, 
see Fig~\ref{fig:state-transition-diagram}.
CT applications are assumed to provide the decoder with SI 
in the form of $\tau_{ij}^{(t)}$, the contact duration,
and $d_{ij}^{(t)}$, a measure of physical proximity.
The infection probability is decided by the CT SI along with individuals’ viral load, $x_{i}$, where only individuals in states infected or infectious have positive viral loads.
More details about our data generation system appear in Appendix~\ref{sec:data_gen_appendix}.

\section{Proposed Group Testing Algorithms}
\label{sec:algos}

This section describes a class of group testing algorithms 
that use side information (SI) within 
generalized approximate message passing~(GAMP)~\cite{rangan2011generalized}
for reconstructing the health status vector $\x$ from the pooled tests, $\y$, given the pooling matrix, $\A$.

\subsection{GAMP-Based Group Testing}
\label{Sec:gamp_binary}
Zhu et al.~\cite{Zhu2020} use GAMP~\cite{rangan2011generalized} for group testing estimation.
GAMP is comprised of two components.
The first component is comprised
of an input channel that relates a prior for $n$ individuals' 
health status, $\x=(x_i)_{i=1}^n$, 
to pseudo data, 
\begin{equation}
\v = \x + \q \in \mathbb{R}^n,
\end{equation}
where the $n$ coordinates of $\x$ are correlated, and
$\q$ is additive white Gaussian noise with $q_i \sim \mathcal{N}(0, \Delta_i)$.
When the $i$th individual is in the infected/infectious state as
defined in Fig.~\ref{fig:state-transition-diagram} and Sec.~\ref{Sec:data_gen},
$x_i = 1$; otherwise, $x_i = 0$.
The pseudo data $\v$ is an internal variable of GAMP that can be considered as a corrupted version of the true unknown health-status vector $\x$. The pseudo data is later iteratively cleaned up to gradually reveal $\x$.
The unknown input $\x$ can be estimated from pseudo data $\v$ using a denoising function (or a denoiser):
\begin{equation}
\widehat{x}_{i}=g_{\text{in}} \left(\v \right) 
= \E \left[X_{i} \mid \V=\v \right] 
\label{input_d},
\end{equation}
where we use the convention that when both the capital and lower case versions of a symbol appear, the capital case is a random variable and the lower case is its realization, and
$\E \left[X_{i} | \v \right]$ represents $\E \left[X_{i} | \V=\v \right]$ when the context is clear.

The second component of GAMP is comprised of an output channel relating the auxiliary vector $\w$ to the noisy measurements $\y$ as reviewed in Sec.~\ref{sec:intro}.
We adopt the output channel {denoiser} of Zhu et al.~\cite{Zhu2020}, 
$h_{j}=g_{\text{out}}\left(y_{j}; k_{j}, \theta_{j}\right) = ( \E\left[W_{j} \mid y_{j}, k_{j},  \theta_{j}\right]-k_{j} ) / \theta_j$, 
where $\theta_{j}$ is the estimated variance of $h_j$, 
and $k_{j}$ is the mean of our estimate for $w_j$. 
Since $y_j$ depends probabilistically on $w_j$, we have 
$f \left(w_{j} \mid y_{j}, k_{j}, \theta_{j}\right) \propto \operatorname{Pr}\left(y_{j} \mid w_{j}\right) \, 
\exp \left[-\frac{\left(w_{j}-k_{j}\right)^{2}}{2 \theta_{j}}\right]$,
where $f$ is a probability density function and $W_j$ is approximated as Gaussian per the central limit theorem when there are enough ones in the $j$th row of pooling matrix $\A$.
Our pseudocode for the GAMP framework is given in Algorithm~\ref{algo:gamp}.
\begin{algorithm}[t]
\caption{Pseudocode for the GAMP Framework}\label{algo:gamp}

{\bf Inputs.} Maximum iterations $t_{\text{max}}$,
side information, probabilities of erroneous tests, measurements $\y$,
and matrix $\boldsymbol{A}$. \\
{\bf Initialize.}
$t,k,h_j,\theta_j,\widehat{x}_i,s_i  \quad \forall i \in [1,n],\ j \in [1,m] $.\\
{\bf Comment.}
$t$ is iteration number,
$k$ is mean of our estimate for $\boldsymbol{Ax}$,
$h_j$ is correction term for $w_j$,
$\theta_j$ is variance of $h_j$,
$\widehat{x}_i$ is our estimate for $x_i$,
$s_i$ is variance in our estimate $\widehat{x}_i$, 
and  $\circ$ is the Hadamard/elementwise product.
\begin{algorithmic}[1]
\While{$t < t_{\text{max}}$}
\State // {\bf clean up the output channel:}
\State $\boldsymbol{\theta}= \A^{\circ 2} \, \s$  // variance of $\h$
\State $\boldsymbol{k}=\A\widehat{\x} - \boldsymbol{\theta} \circ \h$  // mean of $\w$ per previous iteration
\For{$j=1$ to $m$}
\State $h_j= g_{\text{out}}(y_j; k_j, \theta_j)$
\State $r_j= -\frac{\partial}{\partial k_j} g_{\text{out}}(y_j; k_j, \theta_j)$
\EndFor
\State $\boldsymbol{\Delta} = \left[ \frac{1}{N} (\A^{\circ 2})^T \boldsymbol{r} \right]^{\circ -1}$  // scalar channel noise variance
\State $\v = \widehat{\x} + \boldsymbol{\Delta} \circ \A^T \h$  // pseudo data \label{line:q}
\State // {\bf clean up the input channel:}
\For{$i=1$ to $n$}
\State $\widehat{x}_{i}=g_{\text{in}}(\v) = \E[X_i|\v]$  // conditional mean 
\State $s_i = \E[X_i^2|\v] - \E^2[X_i|\v]$  // conditional variance
\EndFor
\State $t=t+1$
\EndWhile
\end{algorithmic}
{\bf Output.} Estimate $\widehat{\x}$, pseudo data $\v$,
and scalar channel noise variance $\boldsymbol{\Delta}$.
\end{algorithm}
GAMP's signal estimation quality is
asymptotically optimal for large reconstruction problems,
and it is also practically useful for finite problem sizes. 
Indeed, numerical results in Sec.~\ref{sec:results} demonstrate 
that GAMP works adequately with a population $n=500$
and well for $n=1000$.

The key to an efficient GAMP-based estimator  
is the input-channel denoiser.
While Zhu et al.~\cite{Zhu2020} considered Bernoulli elements in $\x$
and alluded to supporting non-i.i.d.\ structure in $\x$ by using
vector denoisers,
this paper provides details for probabilistic dependencies within $\x$.
Below, we provide details for the design of two denoisers.
Our first denoiser is based on a probabilistic model that
considers groups of people, such as family members.
Our second denoiser encodes the CT information into the prior for 
each individual's health status.

\subsection{Family Denoiser}
We formalize a family-based infection mechanism,
leading to a denoiser that improves detection accuracy.
Define $\M_{\F}$ as the set of indices of all members of family $\F$.
We say that $\F$ is {\em viral} when there exists viral material in the family.
Next, define the infection probability of individual $i$ within a viral family 
$\F$ as $\pip = \Pr(X_i = 1 \mid \Fviral)$, for all $i \in \M_{\F}$, and $\pif = \Pr(\Fviral)$.
Note that the infection 
status of individuals in a viral family are conditionally 
independent and identically distributed (i.i.d.).

Under our definition, family $\F$ being viral need not be attributed to any individual $i \in \M_{\F}$.
After all, viral material can be on an infected pet or contaminated surface. 
For this model, once the family is viral, the virus
spreads independently with a fixed probability $\pip$.
Of course, our simplified model may not accurately reflect reality. 
That said, without a consensus in the literature on how coronavirus or other infectious diseases spread, 
it is unrealistic to create a more accurate model.
On the other hand, our model is plausible, and we will see that it is mathematically tractable.
We further assume that individuals cannot be infected unless the family is viral,
i.e., $\Pr(X_i = 1 \mid \F \text{ not viral}) = 0$.
The family structure serves as SI and allows the group testing algorithm to impose the constraint that people living together have strongly correlated health status.

Next, we derive the denoiser \eqref{input_d} by incorporating the family-based infection mechanism.
Denote the pseudodata of the members of family $\F$ as $\v_{\F} = (v_i)_{i \in \M_{\F}}$, 
the family-based denoiser for $i$th individual can be decomposed as follows:
\begin{subequations}
\begin{align}
&g_{\text{in}}^\text{family}(\v_\F) \notag \\
=& \E \left[ X_{i} \mid \v_{\F} \right] 
= \Pr(X_i = 1 \mid \v_{\F}) \\ 
=& \Pr(X_i = 1, \Fviral \mid \v_{\F})  \\
=& \Pr(\Fviral \mid \v_{\F}) \Pr(X_i = 1 \mid \v_{\F},\Fviral),
\label{f_den}
\end{align}
\label{eq:denoiser-family}
\end{subequations}%
where the first term of \eqref{f_den} is
\begin{align}
&\Pr(\Fviral \mid \v_{\F}) \nonumber \\
=& \, \frac{f(\v_{\F},\Fviral)}{f(\v_{\F},\Fviral) + f(\v_{\F},\F\text{ not viral}) }.
\label{p_infect}
\end{align}
The two quantities in \eqref{p_infect} can be further expanded as
\begin{subequations}
\begin{align}
&f(\v_{\F},\F\text{ not viral}) \\
= &(1-\pih) \, f(\v_{\F} \mid \F\text{ not viral}) \\
= &(1-\pih) \prod_{i\in \M_{\F}}
\normaldensity{v_i}{0}{\Delta}
\end{align}
\end{subequations}
and
\begin{subequations}
\begin{align}
&f(\v_{\F},\ \Fviral) = \pih \, f(\v_{\F} \mid \Fviral)\\
\begin{split}
  =\, &\pih\sum_{\x_k \in \Omega_{\F}} \prod_{i \in \M_{\F}} \\ 
  & \Big[ 
  f(v_{i} \mid X_{i}=x_{k,i})
  \Pr( X_i = x_{k,i} \mid \Fviral ) 
  \Big],
\end{split}
\label{sum over prod}
\normalsize
\end{align}
\end{subequations}
where $\normaldensity{x}{\mu}{\sigma^2} := \frac{1}{\sqrt{2 \pi \sigma^2}} \exp \left(  -\frac{(x-\mu)^2}{2 \sigma^2} \right)$, 
and $\Omega_{\F} = \{0...00,\  0...10,\ \dots,\ 1...11\}$ is a power set 
comprised of $2^{|\M_{\F}|}$ distinct infection patterns for family $\F$.
The second term of \eqref{f_den} can be simplified as follows:
\begin{subequations}
\begin{align}
\notag
&\Pr(X_i = 1 \mid \v_{\F},\Fviral) \\
= & \Pr(X_i = 1 \mid v_i,\Fviral)  \\
=&\Pr(X_i = 1, v_i \mid \Fviral) \, / \, \Pr(v_i \mid \Fviral) \\
=&\dfrac{\pip \, \normaldensity{v_i}{1}{\Delta}}
{\pip \, \normaldensity{v_i}{1}{\Delta} +  \left(1-\pip\right) \, \normaldensity{v_i}{0}{\Delta}} \\
=&  \left( 1 + \frac{1-\pip}{\pip} \cdot \frac{\normaldensity{v_i}{0}{\Delta}}{\normaldensity{v_i}{1}{\Delta}} \right)^{-1} \\
=& \left( 1 + \left( \pip^{-1} - 1 \right)
  \exp \Big[ \big(v_i-\tfrac{1}{2}\big) \big/ \Delta \Big] \right)^{-1}.
\end{align}
\end{subequations}

\subsection{Contact Tracing Denoiser}
While family structure SI characterizes part of the spread of the disease,
individual family members will presumably all come in close contact with each other;
hence CT SI will include cliques for these individuals. Additionally, CT SI
describes inter-family contacts. Therefore,
CT SI can characterize the spread of the disease more comprehensively than family SI.

Consider a hypothetical widespread testing program that relies on CT SI,
where all individuals are tested $8$ days before the 
testing program begins, resulting in a good initial estimate of 
their ground-truth health status.
Note that our 8 day startup testing assumption 
is in line with China's zero-Covid policy of testing large populations daily for weeks at a time~\cite{Yu-Hsiang}.
To exploit the CT SI, we encode it for each individual $i$ into the prior probability of infection, $\Pr(X_i=1)$,
and use the following scalar denoiser:
\begin{subequations}
\begin{align}
&g_{\text{in}}^\text{CT}(v_i) \nonumber \\
=& \E \left[ X_i \mid v_i \right] = \Pr \left( X_i = 1 \mid v_i \right) \\
=& f(v_i \mid X_i=1) \Pr(X_i=1) / f(v_i) \\
=& \left\{ 1 \!+\! \big[\Pr(X_i\!=\!1)^{-1} \! - \! 1\big]
  \exp \Big[ \big(v_i-\tfrac{1}{2}\big) \big/ \Delta \Big] \!\right\}^{-1}.
\end{align}
\label{eq:denoiser-ct}
\end{subequations}
\hspace{-0.8mm}Here, $\Pr(X_i\!=\!1)$ for day $k+1$ can be estimated by aggregating CT information of individual $i$ over a 
so-called \emph{SI period}\footnote{
Owing to our 8 day startup testing assumption,
starting the SI period on day $k-7$ and ending on day $k$
lets us test on day $k+1$.
Using this SI period
for testing on day $k+1$ implicitly assumes that the test results are available within $24$ hours. However, in cases when it takes more than $24$ hours to generate test results, the SI period can be appropriately modified. For example, day $k-8$ to day $k-1$ for testing on day $k+1$ if it takes at most $48$ hours to generate test results.
\textcolor{red}{After the testing program begins, we assume weekly testing, and probability estimates from the previous group test result $7$ days before the current test 
are used as priors for the $n$ individuals when performing the current group test.}} from day $k-7$ to day $k$ as follows
\begin{equation}
\widehat{\Pr}^{(k+1)}(X_{i}=1) = 1 - \prod_{d=k-7}^k{\prod_{j=1}^n{{ \!\big( 1-\widehat{p}^{(d)}_{i,j} \big) }}},
\label{eq:SI_period_aggre}
\end{equation}
where $\widehat{p}_{i,j}^{(d)}$ is the estimated probability of infection of individual $i$ due to 
contact with individual $j$.
This probability, $\widehat{p}_{i,j}^{(d)}$, 
may be determined by the CT information ($\tau_{ij}^{(d)}, d_{ij}^{(d)})$, as well as their infection status as follows:
\begin{equation}
\widehat{p}_{i,j}^{(d)}=\exp\left(-\big(\lambda \, \tau_{ij}^{(d)} \,  d_{ij}^{(d)} \, \Psi_{ij}^{(d)}+\epsilon\big)^{-1}\right),
\label{eq:est_pij_2nd}
\end{equation}
where $\Psi_{ij}^{(d)} =
1 - \widehat{\Pr}^{(d)}\!(X_i\!=\!0) \, \widehat{\Pr}^{(d)}\!(X_j\!=\!0){\color{orange}}$,
$\lambda$ is an unknown Poisson rate parameter,
and $\epsilon$ is used to avoid division by zero. 
We estimate $\lambda$ with maximum likelihood (ML) using the pseudodata of all individuals, i.e.,
\begin{equation}
\widehat{\lambda}^{\text{ML}} = \arg\max_{\lambda}\ \prod_{i=1}^n f(v_i|\lambda)
\label{eq:mle},
\end{equation}
where \(f(v_i|\lambda) = f(v_i|X_{i}=1) \,  \Pr(X_{i}=1|\lambda)+f(v_i|X_{i}=0) \, \Pr(X_{i}=0|\lambda)\).
Once $\widehat{\lambda}^{\text{ML}}$ is obtained, it is plugged into \eqref{eq:est_pij_2nd} for calculating the prior probability in \eqref{eq:SI_period_aggre}~\cite{dror_plugin}.
As long as the pseudodata is not too noisy, the ML estimated parameter will be close to the true one, and the estimated probability $\widehat{p}_{i,j}^{(d)}$ in \eqref{eq:est_pij_2nd} will be close to the true probability.
This plug-in strategy can also used for our family denoiser $g_{\text{in}}^\text{family}(\v)$,  
where $\lambda = (\pih, \pip)$.

\begin{figure*}[!t]
    \centering\hspace*{-1mm}
    \includegraphics[width=\linewidth]{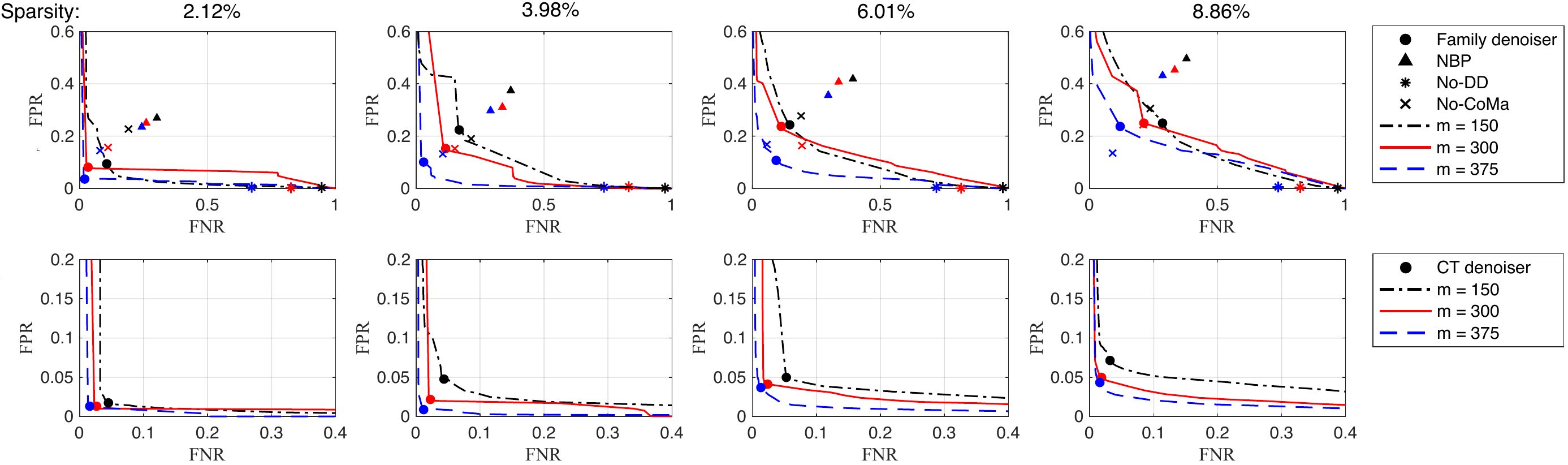}
    \caption{Performance of GAMP-SI in terms of ROC when the family denoiser (top row) and CT denoiser (bottom row) are used. Results for No-DD~\cite{aldridge2019group}, No-CoMa~\cite{CoMa}, and NBP~\cite{bickson2011fault} are included for comparison in the top row. Columns correspond to averaged sparsity levels ranging from $2.12\%$ to $8.86\%$. Within each plot, the performance under three measurement levels for a population of $n = 1000$ individuals is compared. The circular dot on each GAMP-SI curve corresponds to an operating point that minimizes the sum of FPR and FNR. The CT denoiser significantly outperforms the family denoiser with error rates mostly below $0.05$. The estimation problem is more challenging when fewer measurements are used at a higher sparsity level. The results of the baseline algorithm (NBP) that does not exploit SI are worse than both GAMP-based algorithms, since their operating points are far from the ROC curves for the family denoiser; No-DD minimizes FPR at the expense of large FNR; and No-CoMa performs better than GAMP-family at the sparsity level $8.86\%$.} 
    \label{fig:roc_M1}
\end{figure*}

\section{Numerical Results}
\label{sec:results}

\subsection{Experimental Conditions} 
The data were simulated based on the data generation 
process described in Sec.~\ref{Sec:data_gen},
and group testing inference was performed using the algorithms proposed in Sec.~\ref{sec:algos}. We call our family of algorithms GAMP-SI.

\subsubsection{Experiment for n = 1000}
We generated datasets with $n=1000$ individuals using four levels of cross-clique contacts, leading to four averaged sparsity levels, $2.12\%$, $3.98\%$, $6.01\%$, and $8.86\%$, for $\boldsymbol{x}$.\footnote{Because the data generation model was used, it is difficult to control the average sparsity level from the underlying parameters. The sparsity level is measured after the data generation.}
At each sparsity level, we perform pooling experiments using Kirkman triple matrices
as proposed in \cite{Ghosh2021} using $m \in \{150, 300, 375\}$.
Measurement vectors $\y$ for GAMP-SI
were generated using probabilities for erroneous binary tests, $\Pr(y_i=1 \mid w_i=0)=0.001$ and $\Pr(y_i=0 \mid w_i>0)=0.02$, per Hanel and Thurner~\cite{hanel2020boosting}.

\subsubsection{Experiment for n = 500}
We also generated datasets with $n=500$  individuals with averaged sparsity level $1.09\%$. Measurement vectors $\y$ for GAMP-SI were generated under the symmetric noise model, $\Pr(y_i=1 \mid w_i=0)=\Pr(y_i=0 \mid w_i>0)=0.01$,
and $6$ different pooling channel-coded matrices ($m \in \{100, 150, 200, 250,300,350\}$) were derived using a matrix design 
algorithm from Goenka et al.~\cite{Ritesh_G_journal}.

\subsection{Main Numerical Results} 
\label{subsec:results_M1}

We tested our model using both the family denoiser~\eqref{eq:denoiser-family} and the CT denoiser~\eqref{eq:denoiser-ct}.\footnote{We have not included numerical results for the vanilla GAMP implementation without contact tracing SI, because we noticed some numerical instability in the challenging scenarios that involve low sparsity and low measurement rates. Of course, these are the interesting scenarios that group testing focuses on.}
The top row of Fig.~\ref{fig:res-m1-m2} summarizes the performance of GAMP-SI
by choosing a representative operating point on an ROC curve, and Fig.~\ref{fig:roc_M1} presents complete 
receiver operating characteristic (ROC)
curves.
Fig.~\ref{fig:res-m1-m2} reveals that the CT denoiser outperforms the family denoiser in all settings.
Both algorithms yield lower (better) FNR and FPR as the number of measurements, $m$, increases.
Moreover, the CT denoiser's error rates are below $0.05$, except for the challenging cases where the sparsity level is $8.86\%$ and $m\in\{150,300\}$.

Fig.~\ref{fig:roc_M1} illustrates the performance of family and CT denoisers  at different measurement and sparsity levels.
The circular dot on each curve is the operating point that minimizes the total error rate, i.e., the sum of FPR and FNR, which correspond to the concise results of Fig.~\ref{fig:res-m1-m2}.
The closer a dot is to the origin of the FPR--FNR plane, the better the performance it reflects.
Comparing the ROC curves in the top and bottom rows, we note that the CT denoiser significantly outperforms the family denoiser at all sparsity levels.
The CT denoiser, with most of its FNR and FPR $< 5\%$, can achieve as low as $15\%$ of the total error rate of the family denoiser.
Across different sparsity levels, the algorithm performs less accurately as the sparsity level increases. 
In each plot, lower measurement rates make it more challenging for the group testing algorithm.

We also examine the stability of the thresholds corresponding to the operating points we selected to report results in Fig.~\ref{fig:res-m1-m2}. 
Our empirical results reveal that at a given sparsity level, the variation of the threshold due to different design matrices or denoisers is less than $0.003$.
As the sparsity level increases from $2.12\%$ to $8.86\%$, the threshold only drops from $0.160$ to $0.137$.
Hence, the threshold for minimizing the total error rate is insensitive to the testing conditions.

We compare our proposed group testing algorithms to a baseline
nonparametric belief propagation~(NBP) algorithm~\cite{bickson2011fault} that does not exploit SI.
We attach an additional output channel denoiser to the NBP algorithm to process the RT-PCR noise.
We evaluate the performance in terms of FPR--FNR pairs and plot them using triangular markers in Fig.~\ref{fig:roc_M1}. Since the operating points are far from the ROC curves for GAMP using the family denoiser, we conclude that our proposed group testing algorithms that exploit SI 
outperform the NBP baseline that does not use SI.

\textcolor{red}{
GAMP-SI is also compared to the
noisy Column Matching algorithm~(No-CoMa), noisy LP Decoding algorithm~(No-LiPo)~\cite{CoMa}, 
and noisy definite defectives~(No-DD)~\cite{aldridge2019group} 
that do not exploit SI.
No-LiPo and No-DD have large FNR levels, and No-CoMa is tuned for each sparsity level to minimize the sum of FPR and FNR for $m=375$. 
We note that among all algorithms we are comparing with, No-CoMa performs the best and outperforms GAMP-family 
at the sparsity level $8.86 \%$.
We believe the lower performance of GAMP-family at $8.86\%$ is caused by the mediocre SI provided as priors by the weekly group testing regime;
we also observed that GAMP-family suffers from some numerical instability.
In contrast to No-LiPo and No-DD having minimal FPR, 
GAMP-SI can offer any trade-off along the ROC curve.}

\begin{figure}[!t]
    \centering
    \includegraphics[width=\linewidth]{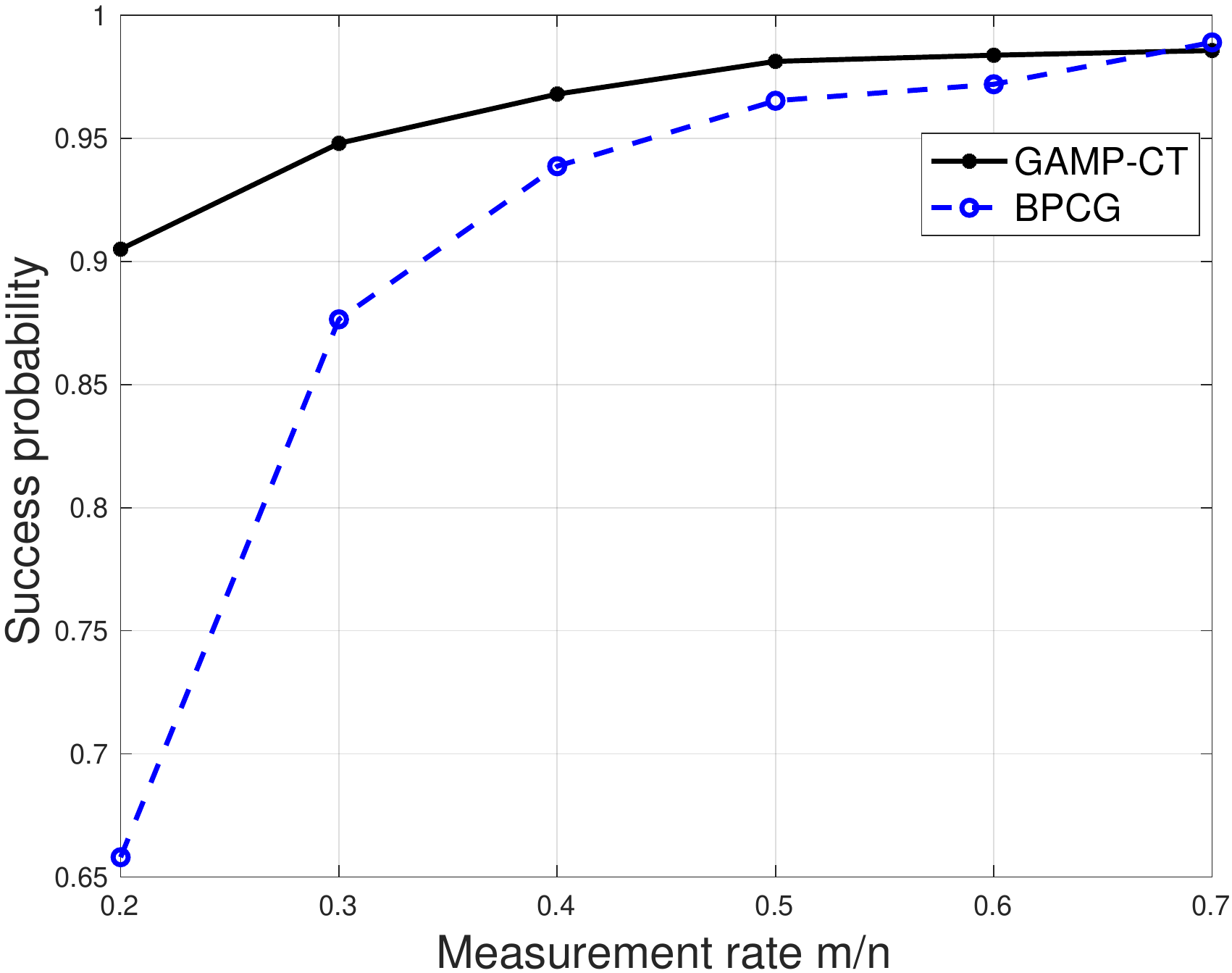}
    \vspace{-2mm}
    \caption{Success probability for algorithms BPCG~\cite{BPCG} and GAMP-SI as a function of the measurement rate $m/n$. With the measurement rate less than $65\%$, GAMP-SI can achieve a higher success probability compared to BPCG. The performance gap is larger when fewer tests are allowed. (Note that BPCG was slow, hence we limited ourselves to $n=500$. However, GAMP-SI performs somewhat better for $n=1000$.)}
    \label{succ_prob}
\end{figure}

We also compare GAMP-SI to belief propagation on combined graphs~(BPCG)\cite{BPCG} that also employs cross-clique contact tracing as SI.
Fig.~\ref{succ_prob} compares the performance of the two algorithms in an experiment for $n= 500$ individuals.
For BPCG, we set the prevalence rate $p$ to $0.01$, the contagion probability $q$ to $0.1$ and the interaction probability
$\theta$ to $0.008$. The number of tests is set to be 
$ m \in \{100,150,200,250,300,350\}$. 
When $m$ is smaller, GAMP-SI has a higher success probability; this advantage vanishes as $m$ increases.
With $m=50$, GAMP-SI can still achieve a success probability of $0.702$.
Also, GAMP-SI has a lower running time than BPCG. 
Specifically,
for $n = 500$, BPCG takes around 680 seconds to produce the result of simulation of one day, on average,
while GAMP-SI only takes around 16 seconds.
Note that BPCG is designed to estimate samples' states based on the information from the previous day; GAMP-SI can use the CT information from the entire previous SI period (multiple days).

\subsection{Additional Experiments} 
\label{subsec:results_M1_additional}
\subsubsection{Using Prior Knowledge of Infection Status}
\begin{figure}[!t]
  \centering
  \includegraphics[width=\linewidth]{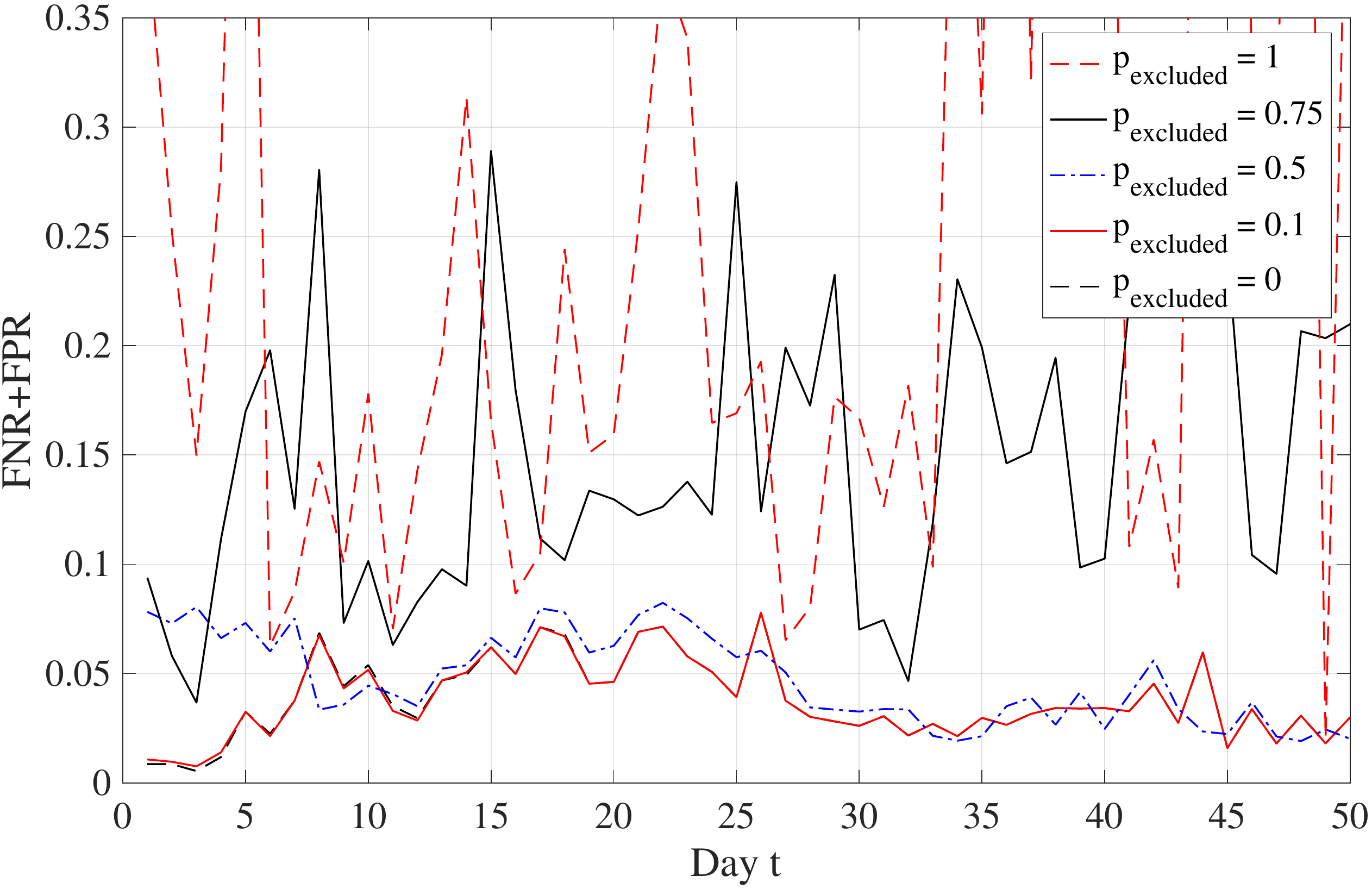}
  \vspace{-2mm}
  \caption{Performance of GAMP-SI when a proportion, $p_{\text{excluded}}$, of the population's health states in the startup phase is unknown.
  The curves reveal that in the absence of up to $50\%$ prior knowledge of the infection status of the population, 
  the accuracy of GAMP-SI is close to that when complete startup information is available.
  }
  \label{fig:edge_due_to_prior}
\end{figure}

We now examine the advantage that prior knowledge of the population's infection status in the startup phase
provides our algorithm for the model.
As defined in \eqref{eq:SI_period_aggre}--\eqref{eq:est_pij_2nd}, 
we iteratively use the updated probability of infection, $\widehat{\Pr}(X_i=1)$, 
estimated from an SI period of 8 immediately
preceding days. 
For days $k < 8$, we had to use the ground-truth infection status of 
each individual in the startup phase to generate the results reported in Sec.~\ref{subsec:results_M1}.
However, ground-truth infection data from 
the startup phase may provide 
our approach an unfair advantage. Below, we investigate whether this advantage is significant.
We examine how varying the amount of startup
information impacts our algorithm's quality.
Specifically, we randomly replace a portion, $p_{\text{excluded}} \in \{0, 0.1, 0.5, 0.75, 1\}$, 
of the population's infection status by an estimated probability of infection, e.g., $5\%$, for a setup that has a true averaged sparsity level of $7.2\%$.
Using a probability instead of a binary value, $0$ or $1$, gives the algorithm soft probabilistic information instead of hard ground-truth style information.
Fig.~\ref{fig:edge_due_to_prior} shows that even with 50\% prior knowledge of the infection status of individuals, 
our detection accuracy for GAMP-SI is close to that when using complete prior information after ramping up for eight days.
The averages of the total error rates across time for increasing $p_{\text{excluded}}$ are $0.038$, $0.039$, $0.046$, $0.148$, and $0.407$, respectively. 
We also tried to replace the startup infection status with an estimated probability of infection of $10\%$, but only observed negligible performance differences.
The results show that
the CT algorithm is robust to the absence of up to $50\%$ of startup infection information, 
suggesting that the startup phase can likely
be optimized to not require onerous testing resources.

\subsubsection{Duration of Startup SI Period}
\begin{figure}[!t]
  \centering
  \includegraphics[width=\linewidth]{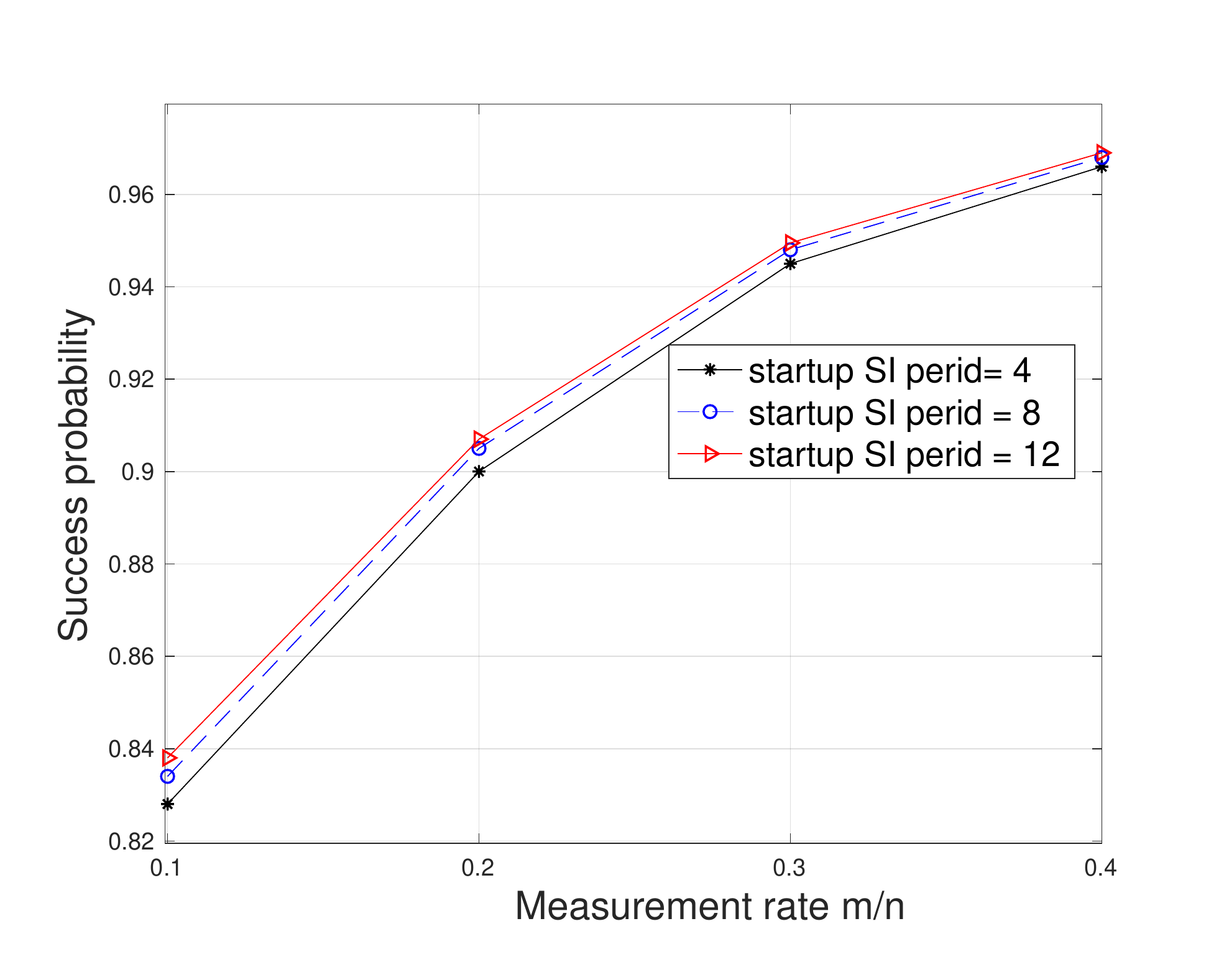}
  \vspace{-4mm}
  \caption{Performance of GAMP-SI when the startup SI period is set to 4, 8, and 12 days.
  The average sparsity level is $2.12\%$ with a sample size of $n = 1000$
  individuals.
  The curves are bunched closely together, suggesting
  that GAMP-SI is not sensitive to the startup SI period.}
  \label{fig:SIperiod}
\end{figure}

It seems plausible to expect
a trade-off between the accuracy of our algorithm and the 
amount of startup infection status data
that needs to be collected before the initialization of 
our testing algorithm. We investigate numerically
whether the accuracy of our algorithm is sensitive to the duration of the startup SI, or the 
startup SI period. We investigated the impact of the startup SI period by testing three 
durations, namely, 4, 8, and 12 days.
The experiment was conducted on a sample of $n = 1000$ individuals with measurement rate $m/n \in \{0.1,0.2,0.3,0.4\}$ at a sparsity level of $2.12\%$.
Fig.~\ref{fig:SIperiod} reveals that the estimation accuracy is not
sensitive to the startup SI period. 
Specifically, as the startup SI period increases from 4 days to 8 days and then 
12 days, the success probability increases by merely
 0.01 and 0.005, respectively, at measurement rate $0.1$. 
 The increase in the success probability is even smaller when the measurement rate becomes larger.
Therefore, we conclude that the startup SI period does not substantially impact the accuracy of 
our algorithm. For this reason, we chose 8
days as the startup SI period for the experiments conducted for this paper.

\section{Discussion} 
\label{sec:discussion}

Recent studies have shown that CT information based on Bluetooth 
may have privacy concerns, and 
contain many errors~\cite{Kleinman2020}. 
However, we note that there exist 
methods for obtaining CT information, which have been shown to be quite effective, including inquiries by social workers, cellphone-based localization, analyses of closed-circuit television footage, and financial transactions~\cite{Hohman2021,Ross2020,CDC_CT}.
A fusion of these modalities can lead to more accurate CT data~\cite{Kleinman2020}. 
Contact tracing has been shown to be useful in previous epidemics~\cite{Kwok2021}. 
As shown earlier, our algorithms are robust to errors in the CT SI. 
However, we leave a full-fledged investigation of SI errors and their impact on our algorithms to future work. We would like to point out that at the peak of a pandemic, testing resources (including time, skilled manpower, testing kits, and reagents) could be scarce.
In such a scenario, it is important to exploit as much information as is reasonably available in order to improve the performance of group testing algorithms, with the aim of saving critical testing resources.

\section{Conclusion} \label{sec:conclusion}

In this paper, we have presented numerical evidence that side information (SI) from family structures and contact-tracing (CT) data can significantly improve the efficiency of group testing.
The overarching message of our paper is that 
using SI within AMP-based approaches 
works well.

There could exist other approaches that also incorporate
SI at the encoder, in addition to the decoder considered 
here, resulting in further reconstruction improvements.
However, we leave a full investigation of this aspect to future work.

Finally, due to the exploratory nature of our work and our deliberate efforts to bring algorithmic advances closer to practice, finding publicly available datasets with associated SI proved to be challenging.
In light of this, we crafted a generative model that closely reflects the key characteristics of COVID-19 transmission for generating the data used in our investigation.
Our numerical results present strong empirical evidence that CT supported group testing is a viable option to optimize the use of resources for a widespread testing program during a pandemic. 
Future work could consider a hybrid approach in which either the contact tracing data or infection status are real, and the other part of the data is simulated. Such an approach would be more realistic without requiring a significant data acquisition project.

\section*{Acknowledgment}
The authors thank Dr. Junan Zhu for allowing them to 
use and modify his implementation of GAMP with SI in their implementation of the family and CT denoisers.
We also thank Karimi et al.~\cite{BPCG} for kindly letting us use their implementation of BPCG.

\appendices

\section{Data Generation Model}
\label{sec:data_gen_appendix}
In this section, we present a generative infection model 
that incorporates CT SI, 
which we use to prepare simulated data for algorithmic evaluation. We model a population of $n$ individuals using a dynamical (time-varying) graphical model that contains nodes, $\{u_i\}_{i=1}^n$, and undirected edges, $\big\{e_{ij}^{(t)}\big\}_{i,j=1}^n$.
On a given day $t$, an edge $e_{ij}^{(t)}$ between nodes $u_i$ and $u_j$ encodes CT SI 
$\big(\tau^{(t)}_{ij}, d^{(t)}_{ij}\big)$, which can be acquired via Bluetooth-based CT applications~\cite{Hekmati2020}. 
Here, $\tau^{(t)}_{ij}$ represents the contact duration and $d^{(t)}_{ij}$ represents a measure of the physical proximity between 
individuals $i$ and $j$.
On day~$t$, a node can be in one of the following states: \emph{susceptible}, \emph{infected}, \emph{infectious}, and \emph{recovered}.
Note that the infected state is defined in a narrow sense that excludes the infectious state, because states must be mutually exclusive.
To keep the model simple, we assume that there are no reinfections, i.e., recovered is a terminal state, despite some reinfections~\cite{Haseltine2020}.
While our model is inspired by a classical compartmental 
model in epidemiology comprised
of susceptible, exposed, infectious, and recovered~(SEIR) states considered for COVID-19~\cite{Carcione2020}, 
our state transitions explicitly use CT
SI and knowledge about the pandemic~\cite{WHOreport}.

We adopt a simplified infection dynamic wherein the
infectious period is preceded and followed by the infected state.
Our design parameters for the infection dynamics are based on a WHO report on COVID-19~\cite{WHOreport}.
Specifically, a node $u_i$ remains infected but noninfectious for $k_1 = 3$ days.
On day $t+k_1$, the node becomes infectious and may transmit the disease to a susceptible neighboring node $u_j$ 
with probability $p_{i,j}^{(t+k_1)}$ whose construction is described below.
An infectious node can potentially transmit the infection until $k_2 = 7$ days after getting infected, and becomes noninfectious afterward. Note that the above assumptions ensure that the transmission of infection on any given day is limited to one hop in the CT graph, i.e., it cannot be the case that individual $X$ infects individual $Y$, who in turn infects individual $Z$ on the same day.
We also model a small fraction of stray infections that may occur, for example, due to sporadic contact with contaminated surfaces. 
Such infections only affect nodes in the susceptible state with a probability $p_1 = 2 \times 10^{-4}$
of our choice.

A state diagram appears in Fig.~\ref{fig:state-transition-diagram}. 
Regarding the viral load $x_i^{(t)}$ for node $i$ on day $t$, we assume $x_i^{(t)} = 0$ if the node is susceptible or recovered.
For an infected or infectious node, we make a simplified assumption that its viral load $x_i^{(t)} \sim \textrm{Uniform}(1, 2^{15})$,\footnote{
The cycle threshold for RT-PCR commonly ranges from $19$ to $34$ cycles \cite[Fig. 3]{Buchan2020}, where $34$ cycles correspond to a low initial viral load of a few molecules, and each cycle roughly doubles the viral density.
Therefore, we estimate the largest possible viral load as $2^{34-19} = 2^{15}$.}
once drawn, remains constant throughout the combined $14$-day period of infection.

Next, we model the probability $p_{i,j}^{(t)}$ that the disease is transmitted from node $u_i$ to $u_j$ on day $t$.
We view infection times as a nonhomogeneous Poisson process with a time-varying rate function $\lambda(t)$.
Consider a $\tau^{(t)}_{ij}$-hour contact on day $t$ when susceptible node $u_j$ is exposed to infectious node $u_i$. 
The average infection rate $\lambda_{ij}(t)$ for day $t$ is assumed to be proportional to both the viral load $x^{(t)}_i$ and the physical proximity $d^{(t)}_{ij}$, namely, $\lambda_{ij}(t) = \lambda_0 \, x^{(t)}_i \, d^{(t)}_{ij}$, where $\lambda_0$ is a tunable, baseline Poisson rate.
The probability that $u_j$ is infected by the end of contact period $\tau^{(t)}_{ij}$ is therefore $p^{(t)}_{i,j} = 1 - \exp \left( -\lambda_0 \, x^{(t)}_i \, d^{(t)}_{ij} \, \tau^{(t)}_{ij} \right)$ for $t \in [k_1, k_2] + t_i$.
From the standpoint of susceptible node $u_j$, all its neighbors $u_k$ that are infectious contribute to its probability of getting infected on day~$t$, namely, $1-\prod_{k} \big( 1 - p^{(t)}_{k,j} \big)$. We remark that if an individual catches the infection on day $t$, then it will possibly be detected only in testing conducted on the day $t+1$, since we assume that sample collection and pooled testing is performed at the beginning of the day.

\bibliographystyle{IEEEtran}
\typeout{}  

\bibliography{IEEEabrv,refs}
\vskip 0pt plus -1fil
\begin{IEEEbiography}[{\includegraphics[width=1in,height=1.25in,clip,keepaspectratio]{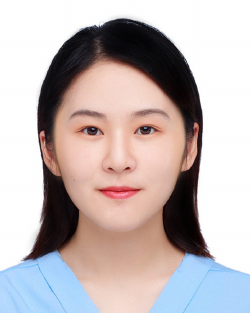}}]{Shu-Jie Cao} received her B.Eng. degree in electronic and information engineering from ShanghaiTech University in 2021. She is currently a Ph.D. student in electrical engineering at Northwestern University. Her research interests include blockchain technologies, signal processing, information theory, and communication networks. Her current projects concern blockchain protocol analysis.
\end{IEEEbiography}

\vskip 0pt plus -1fil
\begin{IEEEbiography}[{\includegraphics[width=1in,height=1.25in,clip,keepaspectratio]{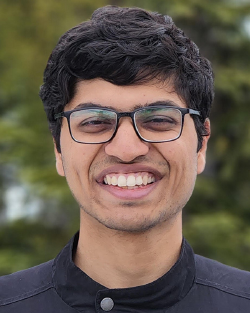}}]{Ritesh Goenka} received a B.Tech. in computer science and engineering from the Indian Institute of Technology Bombay, India, in 2020, and an M.Sc. in mathematics from the University of British Columbia, Canada, in 2023. He will begin a DPhil in mathematics at the University of Oxford in October 2023. His research interests span discrete mathematics, including combinatorics, discrete probability, and number theory.
\end{IEEEbiography}

\vskip 0pt plus -1fil
\begin{IEEEbiography}[{\includegraphics[width=1in,height=1.25in,clip,keepaspectratio]{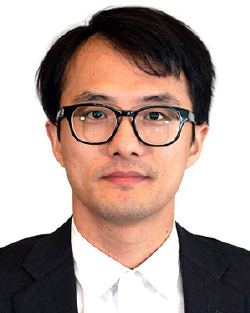}}]{Chau-Wai Wong}
(Member, IEEE) received his B.Eng. and M.Phil. degrees in electronic and information engineering from The Hong Kong Polytechnic University, in 2008 and 2010, and his Ph.D. degree in electrical engineering from the University of Maryland, College Park, MD, USA, in 2017. He is currently an Assistant Professor with the Department of Electrical and Computer Engineering, the Forensic Sciences Cluster, and the Secure Computing Institute, at North Carolina State University. He was a data scientist at Origin Wireless, Inc., Greenbelt, MD, USA. His research interests include multimedia forensics, statistical signal processing, machine learning, data analytics, and video coding. Dr. Wong received a Top-Four Student Paper Award, Future Faculty Fellowship, HSBC Scholarship, and Hitachi Scholarship. He was the General Secretary of the IEEE PolyU Student Branch from 2006 to 2007. He was involved in organizing the third edition of the IEEE Signal Processing Cup in 2016 on electric network frequency forensics.
\end{IEEEbiography}
    
\vskip 0pt plus -1fil
\begin{IEEEbiography}[{\includegraphics[width=1in,height=1.25in,clip,keepaspectratio]{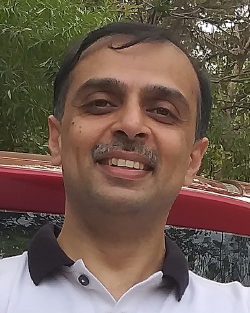}}]{Ajit Rajwade} (Senior Member, IEEE) is an Associate Professor in the Department of Computer Science and Engineering at IIT Bombay. He is also associated with the Center for Machine Intelligence and Data Science (CMInDS) and the Koita Center for Digital Health (KCDH) at IIT Bombay. He is a senior member of the IEEE and an associate editor for the journal Signal Processing (Elsevier). He obtained his bachelor's degree from the University of Pune, India, in 2001, his master's degree from McGill University, Montreal, Canada, in 2004, and his Ph.D. degree from the University of Florida, Gainesville, USA, in 2010, all in computer science and engineering. His research interests are in image and signal processing with a focus on compressed sensing, matrix and tensor recovery, tomographic reconstruction, image restoration and compression, group testing, and probability density estimation. He is a recipient of an excellence in teaching award from IIT Bombay in 2019 and a best scientific paper award from the International Conference on Pattern Recognition in 2008. 
\end{IEEEbiography}

\vskip 0pt plus -1fil
\begin{IEEEbiography}[{\includegraphics[width=1in,height=1.25in,clip,keepaspectratio]{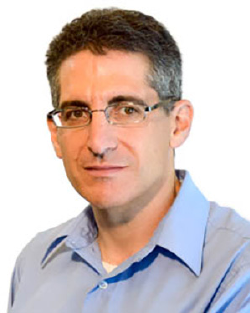}}]{Dror Baron}
(Senior Member, IEEE) received the B.Sc. (summa cum laude) and
M.Sc. degrees from the Technion--Israel Institute of Technology, Haifa, Israel,
in 1997 and 1999, and the Ph.D. degree from the University of Illinois at Urbana-Champaign in 2003, all in electrical engineering.
From 1997 to 1999, Dr. Baron worked at Witcom Ltd. in modem design.
From 1999 to 2003, he was a research assistant at the University of Illinois at
Urbana-Champaign, where he was also a Visiting Assistant Professor in 2003.
From 2003 to 2006, he was a Postdoctoral Research Associate in the Department
of Electrical and Computer Engineering at Rice University, Houston, TX. From
2007 to 2008, he was a quantitative financial analyst with Menta Capital, San
Francisco, CA, and from 2008 to 2010 a Visiting Scientist in the Department
of Electrical Engineering at the Technion--Israel Institute of Technology,
Haifa. Since 2010,  Dr. Baron has been with the Electrical and Computer Engineering
Department at North Carolina State University, where he is currently an
Associate Professor.

Dr. Baron's research interests combine information theory, signal processing,
quantum computing, and fast algorithms.
Dr. Baron was a recipient of the 2002 M. E. Van Valkenburg Graduate Research Award, and received an honorable mention at the Robert Bohrer Memorial Student
Workshop in April 2002, both at the University of Illinois. He also participated
from 1994 to 1997 in the Program for Outstanding Students, comprising the top
0.5\% of undergraduates at the Technion.
\end{IEEEbiography}
\balance{}

\end{document}